\documentclass[twocolumn]{aastex631}

\usepackage{comment}
\usepackage{amsmath,amssymb}
\usepackage{verbatim}
\usepackage{graphicx}
\usepackage{mathrsfs,yfonts}
\usepackage{latexsym}
\usepackage{amsthm}
\usepackage{braket}

\usepackage{amssymb}
\usepackage{hyperref}
\usepackage{color}
\usepackage{footmisc}

\shorttitle{Mad Halos}

\graphicspath{{./}{figures/}}

\newcommand{\gsim}{\lower.7ex\hbox{$\;\stackrel{\textstyle\rangle}{\sim}\;$}}
\newcommand{\lsim}{\lower.7ex\hbox{$\;\stackrel{\textstyle\langle}{\sim}\;$}}

\definecolor{summersky}{cmyk}{0.71,0.33,0,0.14}
\definecolor{flamingo}{cmyk}{0,0.51,0.71,0.14}

\begin{document}

\title{Spin Alignment of Dark Matter Halos: Mad Halos}



\author{Ehsan Ebrahiman}
\affiliation{Sharif University of Technology \\
Azadi Ave.\\
Tehran, Iran}

\author[0000-0002-4442-1523]{Aliakbar Abolhasani}
\affiliation{Sharif University of Technology \\
Azadi Ave.\\
Tehran, Iran}





\begin{abstract}
We investigate the spin alignment of the dark matter halos by considering a mechanism somewhat similar to tidal locking; we dubbed it Tidal Locking Theory (TLT). While Tidal Torque Theory is responsible for the initial angular momentum of the dark matter halos, the Tidal locking Theory explains the angular momentum evolution during non-linear ages. Our previous work showed that close encounters between haloes could drastically change their angular momentum. The current manuscript argues that the tidal locking theory predicts partial alignment between speed and the spin direction for the large high-speed halos.  To examine this prediction, we use the IllustrisTNG simulation and look for the alignment of the halos’ rotation axis. We find that the excess probability of alignment between spin and speed is about 10 percent at $z=0$ for fast haloes—with velocities larger than twice the median. We show that tidal torque theory predicts that the spin of a halo tends to be aligned with the middle eigendirection of the tidal tensor. Moreover, we find that the halos at $z=10$ are preferentially aligned with the middle eigendirection of the tidal tensor with an excess probability of 15 percent. We show that tidal torque theory fails to predict correct alignment at $z=0$ while it works almost flawlessly at $z=10$.\end{abstract}



\section{Introduction} \label{sec:intro}
{Knowing the origin and evolution of the angular momentum of galaxies is essential for both galaxy-evolution theories and galaxy-lensing data \cite{2008MNRAS.391..197B,Merkel:2013tku}. In particular, the galaxy-lensing surveys are sensitive to galaxy alignment \cite{Schaefer:2008xd,Codis:2014awa}. These surveys can nail down a handful of parameters for precision cosmology. On the other hand, the correlation between galaxy angular momentum and the large-scale structure (LSS) of the universe is known in both data \cite{Zhang:2014rju,Kraljic:2021oeg} and simulations \cite{Chisari:2015qga,2012MNRAS.427.3320C}. Since such correlations could contaminate the lensing data, many efforts have been underway to predict the galaxies' angular momentum alignments theoretically \cite{Codis:2015tla,Laigle:2013tsa}. One of the main challenges is that we observe the baryonic matter, while most theoretical models have predictions for the dark matter halos. Namely, the relation between the angular momentum of the galaxy and the dark matter halo is unclear \cite{Jiang:2018ioo}. Moreover, today we know that the spin direction of the dark matter halos deviates from the predictions \cite{Lopez:2020adh}. We will mainly focus on the latter problem in this work.}

Tidal Torque Theory (TTT) has a viable well-settled explanation for the origin of the angular momentum. Hoyle suggested that when the matter collapses to form a galaxy/halo, the background gravitational tidal field applies a torque on the infalling matter that changes its angular momentum \cite{1951pca..conf..195H}. Peebles estimated the net angular momentum acquired by a collapsing spherical overdense region \cite{1969ApJ...155..393P}. He used the linear perturbation theory to estimate angular momentum acquisition during the linear stage. Dorshkevich and later White pointed out that assuming spherical symmetry for the collapsing matter is unnecessary and perhaps misleading \cite{1970Afz.....6..581D,White:1984uf}. They argue that the primary explanation for the angular momentum of halos is the misalignment between the tidal tensor of the ambient matter and the inertial tensor of the infalling matter that exerts a torque on the infalling matter.

To be more precise, take all the particles that end up in a single halo, and  by using the Zeldovich approximation \cite{Zeldovich:1969sb}, write the physical position of them, $\boldsymbol{r}$, in terms of the Lagrangian comoving coordinate $\boldsymbol{q}$, and The Newtonian potential,$\Phi$ :

\begin{equation}
    \boldsymbol{r}=a(t)[\boldsymbol{q}-D_{+}(t)\boldsymbol{\nabla}\Phi],
\end{equation}
{where $D_{+}(t)$ is the linear growth function, the above equation holds when the matter perturbation is linear, and the dark matter halo has not formed, i.e., there is only a proto-halo. Assuming all dark matter particles that will end up in a final halo are inside a Lagrangian volume, $V_L$, the angular momentum of these particles relative to the center of mass, $\int d^3\boldsymbol{r}\,\rho\, \boldsymbol{r}\times\boldsymbol{v}$  is found to be}
\begin{equation}
    \boldsymbol{L}(t)=-a^5\Dot{D}_+ \int_{V_L}d^3\boldsymbol{q}\,(\boldsymbol{q}-\boldsymbol{q}_{cm})\times\boldsymbol{\nabla}\Phi.
\end{equation}
Expanding the potential around the center of mass to the second-order, we get
\begin{equation}
    \Phi(\boldsymbol{q})=\Phi(\boldsymbol{q}_{cm})+\Delta q^i\frac{\partial\Phi}{\partial q^i}\Bigg|_{q_{cm}}+\Delta q^i\Delta q^j \frac{\partial^2\Phi}{\partial q_i \partial q_j}\Bigg|_{q_{cm}}
\end{equation}
The second term is the acceleration of the center of mass, and the third one is the tidal field. The tidal field is the leading non-vanishing term contributing to the angular momentum-- relative to the center of mass. Therefore, we find
\begin{equation}\label{eq:TTTI}
    L_i(t)=-a^2\Dot{D}_+\,\epsilon_{ijk} \frac{\partial^2\Phi}{\partial q^l \partial q^k}\Bigg|_{q_{cm}} \int_{V_L} d^3\boldsymbol{q}\,\rho(t)a^3 \Delta q^j \Delta q^l.
\end{equation}
In the above equation, the integral is the inertial tensor of the in-falling matter in the Lagrangian coordinate, called $\boldsymbol{I}$; and the potential second derivative, namely tidal tensor, is denoted by $\boldsymbol{T}$ in the rest of the paper. Both $\boldsymbol{I}$ and $\boldsymbol{T}$ are constant in the linear stage. Therefore,  $\boldsymbol{L}$ depends on time only through $a^2 \dot{D}_+$ factor above. In a  matter-dominated universe $a^2\dot{D}_{+}=t$, so $L_i(t)=-t \epsilon_{ijk} I_{jl}T_{lk}$. It means that in a matter-dominated universe, the angular momentum of a proto-halo, growth linearly with time. This linear growth continues until the matter stops expanding, i.e., turnaround time, after which matter collapses to form the halo, and angular momentum stops growing. Hence, at the turnaround time, $t_m$, the final angular momentum is
\begin{equation}\label{eq:TTTL}
    L_i\approx -t_m \epsilon_{ijk} I_{jl}T_{lk}
\end{equation}

Therefore, one can investigate the prediction of TTT in N-body simulations. Many such investigations have been performed since the theory was developed  \cite{White:1984uf,Barnes:1987hu,Sugerman:1999au,Porciani:2001db,Lopez:2018lnz,Zjupa:2016xpk}. These simulations confirm linear growth of the angular momentum of \emph{proto-halo}s. Namely, if one marks all the present-time halo particles and traces them back to the initial state, their angular momentum changes like the TTT prediction, linear with time. Hence it is reasonably precise to say that nothing can compete with the TTT mechanism at the proto-halo or linear stage. However,  TTT has some critical shortcomings.  Many studies have shown that the direction of the angular momentum deviates from the prediction of the TTT \cite{Porciani:2001db,Lopez:2020adh}. Moreover, the TTT is silent about the evolution of the angular momentum after the halo being formed. We know that the torque on a halo-- through the tidal field of the ambient matter-- decreases dramatically when the halo collapses and becomes Virialized. So TTT erroneously claims that after the formation, the angular momentum remains constant.
Some studies tried to explain the angular momentum evolution by taking mergers into account \cite{Vitvitska:2001vw,Bett:2015aoa,Drakos:2019tel,Benson_2020}. However, another study states angular momentum evolution can be solely predicted by knowing the initial conditions \cite{Cadiou:2020rdj}; since the same tidal field is responsible for accelerating the merging halos, it is not surprising that the initial conditions completely determine the final angular momentum. In other words, mergers can be considered to be precisely like the accretion of matter into halos.  Other studies try to explain the deviations from the TTT by extending the calculations to the non-linear stage \cite{10.1093/mnras/stv549,Laigle:2013tsa,Porciani:2001db}. 
\\
In our previous work \cite{Ebrahimian:2020tmy}, we studied a mechanism responsible for the angular momentum evolution during non-linear stages, Tidal Locking Theory (TLT); and investigated how it affects the angular momentum evolution. According to this theory, if two halos pass by, they exert torque on each other, which changes the angular momentum of each halo. It is somewhat like the mechanism of the tidal locking in our Solar System. This article aims to calculate the angular momentum change according to the TLT theory in more detail to address dark matter halos spin alignment. Finally, we investigate the Illustris simulation and find evidence supporting the TLT mechanism on the alignment of the halos.

\section{Tidal Locking and Alignment}\label{sec:TLT}

Dark matter halos tend to cluster due to gravitational interactions, so their environment is relatively dense at late times, and fly-by of halos are common. In close encounters, two halos can change their angular momentum with a somewhat similar mechanism responsible for the tidal locking of planets and moons. In the case of planetary tidal locking, tidal forces reshape the planet or moon into an ellipsoid. The period of tidal force is the same as the orbital period, so the difference between rotation and orbital period causes the misalignment between the ellipsoid and tidal force. This misalignment leads to a torque that changes the rotational velocity and angular momentum. 
However, in planetary tidal locking, objects are compact, rigid, and initially spherical, but the dark matter halos are extended, diffuse, and non-spherical. Initial overdense regions, generically non-spherical, collapse to form dark matter halos by gravitational instability \cite{Bardeen:1985tr}. Noteworthy that dark matter halos are not necessarily spherical even after Virialization \cite{Allgood:2005eu,Emami:2020cwt}. Also, the large size of the halos increases the torque, so a single encounter is enough to change the angular momentum of the dark matter halos. It must be noticed that since the dark matter halos are not solid, complete locking is impossible, we use this name to mention the similarity between this mechanism and tidal locking in the planetary systems.
\\
Now we are going to calculate the angular momentum change in a single event. We estimated the magnitude of the angular momentum change in such an encounter event in the previous work, and here we mainly focus on the alignment. Let us assume a halo with mass $M_1$ and size $R$; another halo with mass $M_2$ passes by $M_1$ with impact factor $b$, and relative speed is $v$. By the same approximation that led us to \ref{eq:TTTI}, we can write the TLT torque

\begin{equation}
    \tau_{i}=\epsilon_{ijk} I_{jl}\partial_l\partial_k \Phi.
\end{equation}
$\boldsymbol{I}$ is the inertial tensor and $\partial_i\partial_j\Phi$ is the tidal tensor, both are computed in the physical coordinates rather than the Lagrantian coordinates, and $\Phi$ is the gravitational potential with respect to the $M_2$ at $M_1$. It must be noted that $\Phi$ in Eq.4 is the gravitational potential of the ambient matter, while The passing halo sources $\Phi$ in the above equation. To find the angular momentum change of a halo, we integrate the torque exerted on a halo, so 
\begin{equation}\label{eq:TLT1}
    L_{i}=\epsilon_{ijk} I_{jl} \mathcal{M}_{lk}
\end{equation}
where we defined $\mathcal{M}_{lk}=\displaystyle \int dt\, \partial_l\partial_k \Phi$.  Let us assume $v^2\gg G(M_1+M_2)/b$, so halos traverse a linear path to a very good approximation. In the following, we will discuss more the importance of containing to these halos. By this assumption we find that 
\begin{equation}\label{eq:TLTM}
    \mathcal{M}_{ij}=\frac{2GM_2}{b^2v} (\hat{b}_i\hat{b}_j-\hat{c}_i\hat{c}_j)
\end{equation}
where $\boldsymbol{\hat{b}}$ is the unit vector on the orbital plane  perpendicular to the $\boldsymbol{v}$ ; and $\boldsymbol{\hat{c}}$ is the unit vector perpendicular to the orbital plane.

What so far we have found is the angular momentum change for a single encounter event. However, to compute the angular momentum itself, we must know the $I_{ij}$ for each halo, which is not straightforward even for the simulations. To overcome this ignorance, we need to average over different halo shapes and configurations. With halo shapes being independent of the environment, $\langle I_{ij} \rangle=\delta_{ij}$, which results in $\langle L_i \rangle=0$. However, it must be noted that this does not mean that there is no preferred orientation of the angular momentum vector. Let us introduce a indicator for alignment. Assume an ensemble of unit vectors, $\{\hat{\boldsymbol{u}}^k\}$, that we wish to find their alignment. If one wants to find out that the $\{\hat{\boldsymbol{u}}^k\}$ is aligned with a specific unit vector-like $\hat{\boldsymbol{v}}$, a common way is to find the distribution function of $\mu=\boldsymbol{\hat{u}.\hat{v}}$. If the $\{\hat{\boldsymbol{u}}^k\}$ is completely random,  the distribution function of $\mu$ must be  uniform. In the data, the distribution function of $\mu$ is symmetric under $\mu\rightarrow -\mu$,for this reason, $\braket{L_i}=0$. Hence, one only needs to find the distribution function of $|\mu|$. Particularly, it must be noted that a random distribution for an ensemble of vectors is $P(|\mu|)=1$.

Hence, one can define the deviation of $P(|\mu|)$ from a uniform distribution to be a suitable measure of alignment. In particular, we use $\xi=P(|\mu|)-1$ as a viable measure of alignment. This measure have been used in \cite{2012MNRAS.427.3320C} and known as \emph{excess probability of alignment.} Note that if $\xi$ increases with $\mu$, then $\{\hat{\boldsymbol{u}}^k\}$ tends to be aligned with $\hat{\boldsymbol{v}}$ and vice versa. The more substantial alignment leads to bigger $\xi$. Nevertheless, it is not clear that for a given ensemble of vectors like $\{\hat{\boldsymbol{u}}^k\}$, we could find the vector $\hat{\boldsymbol{v}}$ with the strongest alignment, in other words, we could find the preferred orientation for a set of vectors.

Consider an ensemble of halos with $N_h$ halos, and $\hat{L}_i(n)$ denotes the spin unit vector of every individual halo; the average of $\hat{L}_i\hat{L}_j$ is
\begin{equation}
A_{ij}=\braket{\hat{L}_i\hat{L}_j}=\frac{1}{N_h}\sum_{n=1}^{N_h} \hat{L}_i(n)\hat{L}_j(n).
\end{equation}
It must be noticed that $A_{ij}$ is a positive-definite matrix with trace $1$, $\mathrm{Tr}[A]=1$.
Let us denote the eigenvalues and eigenvectors of $A$ by $\{\lambda_a ;\hat{J}^a_i\}$; one can show show that $\lambda^a=\braket{(\boldsymbol{\hat{L}.\hat{J}}^a)^2}=\braket{\cos^2\theta_a}$, where $\theta_a$ is the angle between the $\boldsymbol{\hat{L}}$ and $\boldsymbol{\hat{J}}^a$. Now, for an arbitrary vector $\hat{\boldsymbol{v}}$ with unit length, $\braket{(\boldsymbol{\hat{L}.\hat{v}})^2}$ can be found as
\begin{equation}
    \braket{(\boldsymbol{\hat{L}.\hat{v}})^2}=\sum_a \lambda_a \cos^2\alpha_a. 
\end{equation}
where $\cos\alpha_a= \boldsymbol{\hat{v}.\hat{J^a}}$. The statistical orientation of angular is a vector $\hat{\boldsymbol{v}}$ for which the above expression maximizes. It can be shown that the $\hat{\boldsymbol{v}}$, maximizing the above expression, is nothing but the eigenvector of  $A$ with the largest eigenvalue. This argument can be generalized to the angular momentum vector, including its magnitude; therefore, the eigenvector with the largest eigenvalue of $\braket{\hat{L}_i\hat{L}_j}$ determine the angular momentum preferred orientation.

If we use Eq. \ref{eq:TLT1} in an appropriate coordinate system in which the $\mathcal{M}_{ij}$ is diagonalized, say $\mathcal{M}=\mathrm{diag}(\lambda_1,\lambda_2,\lambda_3)$; we find
\begin{equation}\label{eq:Aling1}
\braket{L_i L_j}=B\begin{pmatrix}
(\lambda_2-\lambda_3)^2 & 0& 0\\
 0& (\lambda_1-\lambda_3)^2 &0 \\
 0&0 & (\lambda_1-\lambda_2)^2
\end{pmatrix}
\end{equation}
$\lambda_i$ is the eigenvalue of $\mathcal{M}_{ij}$, and $B=\frac{1}{30}(3\braket{I_{ij}^2}-\braket{I_{ii}^2})$. 

Let us remark here on the TTT. Note that if one assumes that $\mathcal{M}_{ij}=-t_m T_{ij}$, Eq. \ref{eq:Aling1} will reduce to the TTT’s result. Consequently, if TTT were the \emph{only} mechanism involves in spinning  halos, after averaging over shapes, the spin of halos partially aligns with the second eigenvector of the tidal tensor, $\partial_i\partial_j \Phi$.  In other words, the distribution of the spin direction is not entirely isotropic, but it is distributed with a tendency to orient perpendicular to the plane that the eigenvalues $\partial_i\partial_j\Phi$ have the largest difference. This is in agreement with \cite{Lee:1999ii}.

Now let us move on to the TLT theory. Using Eq.\ref{eq:TLTM}, we first perform ensemble-average over different halo shapes. Therefore, for a single encounter event, we get
Now let's return to the TLT theory. Using Eq. \ref{eq:TLTM}, and averaging over shapes, for a single encounter event, we get\begin{equation}\label{eq:TLT2}
    \braket{L_i L_j}=S_{ij}(M_2,b,\boldsymbol{v})=4 B \left(\frac{GM_2}{b^2 v} \right)^2 \big[\delta_{ij}+3\hat{v}_i\hat{v}_j\big],
\end{equation}
where $\boldsymbol{\hat{v}}$ is the unit vector in the direction of velocity. In the above equation, the factor $3$ behind the second term in brackets means that after a single encounter, the probability of  angular momentum change being in the direction of the relative velocity twice higher than in other directions, statistically.

As we mentioned before,  halos usually reside in crowded regions so that a single halo may experience several encounters after its formation.  Each encounter has a different relative velocity and impacts factor. That being the case, the final angular momentum of a single halo is hard to predict. In principle, one can compute the expected value of the $\braket{L_iL_j}$
\begin{align}
\nonumber
\braket{L_i L_j}_{TLT}= &  \int dt dM d^3\boldsymbol{v}d^3\boldsymbol{r}\, S_{ij}(M,|\boldsymbol{r-r_0}|,\boldsymbol{v-v_0})\times \\
     & f(M,\boldsymbol{r},\boldsymbol{v})\delta\left(\frac{\boldsymbol{(r-r_0).(v-v_0)}}{|\boldsymbol{v-v_0}|}\right) |\boldsymbol{v-v_0}|
\end{align}
where $f(M,\boldsymbol{r},\boldsymbol{v})$ is the distribution function of the neighboring halos, for a halo at $\boldsymbol{r_0}$, with the speed $\boldsymbol{v}_0$ . Nevertheless, it is hard to have a general prediction from the above relation. Even so, for a mad halo, namely with $v_0\gg v_{\mathrm{rms}}$, the above integral reduces to the following expression
\begin{equation}
    \braket{L_i L_j}=\int dt\, v_0 \int db\, 2\pi b \, \int dM\, n(M)~ S_{ij}(M,b,\boldsymbol{v}_0), 
\end{equation}
in which $n(M) dM$ denotes the number density of halos with mass between $M$ and $M+dM$. Mad halos have probably undergone a fly-by and acquire their high speed. On the other hand, owing to their high speed, all relative velocities in each encounter are approximately $\boldsymbol{v}_0$, so for such halos, one can conclude that the angular momentum aligns with the velocity of the halos. It must be noticed that for mad halos at each encounter, $S_{ij}\propto [\delta_{ij}+3\hat{v_0}_i\hat{v_0}_j]$, so $[\delta_{ij}+3\hat{v_0}_i\hat{v_0}_j]$ can consequently be factorized.

\section{Data and Analysis}
Nowadays, computer simulations have become an indispensable tool for studying cosmology. In particular, due to the non-linear nature of the large-scale structure's physics, it is almost impossible to rely only on analytical calculations. The IllustrisTNG simulations are a series of large-box cosmological simulations. The results of TNG50 simulation released by The TNG project team in early 2021. The simulation volume is roughly $ 50^3 \mathrm{Mpc}^3$, and the mass resolution is higher than previous runs, which simulate a single dark matter halo more accurately. These characteristics allow us to investigate the small and large structures simultaneously, which is crucial for investigating spin-LSS alignments. Hence, we use TNG50-1-Dark simulation in this study. The halos or Groups are derived with the FoF algorithm, and the smaller structures called subhalos are identified with the Sub-Find algorithm. In this sense, halos are groups of subhalos. Here, we use "halo" for the subhalos and not  the Groups. The data of the speed and angular momentum vectors, mass, and position of the halos are available for each snapshot. We use these data to investigate spin alignments over time and compare them with TTT and TLT predictions. Some other details of the TNG50-Dark-1 simulation are shown in table \ref{tab:simulation}.

\begin{table}[t]
    \centering
    \begin{tabular}{ll}
    \hline
        $L_{\rm box}$ (comoving) & 51.7 Mpc \\
        $m_{DM}$ & $5.4\times 10^5 M_{\odot}$\\
        $N_{DM}$ & $2160^3 $\\
        $N_{\rm snap}$ & 100\\
        $\Omega_{m}$ & 	0.3089\\
        $\Omega_{\Lambda}$ & 0.7274\\
       $ \Omega_{bar}$ & 0\\
        h & 0.6774\\
        $N_{\rm subfind}(z=0)$& 6616159\\
        \hline
    \end{tabular}
    \caption{TNG50-1-Dark simulation}
    \label{tab:simulation}
\end{table}

 We investigate predictions of both the TTT and TLT at redshifts $z=0,1,4,10$. Firstly, we determine the net spin alignment direction, $\hat{\boldsymbol{v}}$; after that, we find the distribution function of $\mu=\boldsymbol{\hat{L}.\hat{v}}$ using the simulation.
 
 \subsection{Tidal Locking Theory}
Sec.\ref{sec:TLT} argued that mad halos, i.e., with high velocity, have spin partially aligned with their speed in a statistical sense. Now we focus on the excess probability of the alignment between velocity and angular momentum vectors of mad halos, $\mu=\cos(\theta)=\boldsymbol{L.v}/|\boldsymbol{v}||\boldsymbol{L}|$. To be concrete, for a mad halo, we define a halo in the simulation which $v>2\braket{v}$ where $\braket{v}$ is the median of the halo velocities.\footnote{The factor 2 is arbitrary, one can define mad halo as $v>a\braket{v}$ by taking $a\gtrsim 1$ }
According to Eq.\eqref{eq:TLT2}, the size of the angular momentum change during a single encounter decreases with speed. One may worry that mad halos are not good candidates to study the TLT. We further explain this point in the following. It is common to define the dimensionless spin parameter
\begin{equation}
    \lambda=\frac{L\sqrt{|E|}}{GM^{5/2}},
\end{equation}
in which $E$ is the energy (kinetic+potential) of the halo, and $M$ is the halo's mass. $\lambda\sim 1$ means that all halo particles are rotating in the same direction, and $\lambda\ll 1$ means that there is no net rotation. Typically the spin parameter has an amplitude between $0.04$ and $0.07$ that does not depend on the mass and other characteristics of the halo \cite{Steinmetz:1994nb,Zjupa:2016xpk}. We use Eq. \eqref{eq:TLT2} to estimate the spin parameter change during one impact; we assume that the halo is virialized, therefore
\begin{equation}
    \lambda_{TLT}\sim \left(\frac{GM_2}{v^2b}\right)^{1/2} \left(\frac{R_1}{b}\right)^{3/2}
\end{equation}
$R_1$ is the size of the halo. So by choosing the large mad halos, we can compensate for their lower TLT effects due to the high speed. The half-mass radius of the halos is available on all IllustrisTNG simulations; hence we take it as the size of the halo.  We will calculate $\mu$ for large mad halos, namely mad halos that $R>2\braket{R}$, which $\braket{R}$ is the median of halos' half-mass radius.

\subsection{Tidal Torque Theory}
Although it is not our primary goal, we check the prediction of the TTT for the spin-alignment to confirm our theoretical predictions. Performing the above calculations for checking the TLT are pretty straightforward, we only need to find the angle between the velocity and the spin of halos, but the calculations for TTT are a bit subtle; first, we should compute the gravitational potential and $\partial_i\partial_j\Phi$ at each point. Then we calculate the eigenvectors of $\partial_i\partial_j\Phi$, take $\hat{\boldsymbol{e}}_1,\hat{\boldsymbol{e}}_2,\hat{\boldsymbol{e}}_3$ as the eigenvectors of the largest, middle and smallest eigenvalue of the tidal tensor. Finally we find the angle between angular momentum and $\hat{\boldsymbol{e}}_2$, and calculate the excess probability by using the simulation.

To do this,  we first divide the simulation box into $n^3$ grids, and then we find the total mass of each grid by summing over the masses of halos within the grid. We take $n \approx \ell/t\braket{v}\approx 15$, which $\ell$ is the size of the simulation box. This discretization helps us to find halos near their formation environment. Noteworthy that the TTT effects are significant at higher redshifts. However, we must wait until halo formation proceeds enough to determine the mass of each grid accurately. Nearly half of the present time subhalos have been formed by the redshift $10$; this is early enough to still look for the effect of the TTT. We think this sign of the TTT begins to fade away after this redshift due to the effects like TLT and moving halos from their initial position.

Having found the mass of each grid, we can calculate the potential on this grid. First, we use the discrete Fourier transform of the mass field and divide it by the $K^2=D_1^2+D_2^2+D_3^2$,  which $D_i=\sinh{2\pi i k_i/n}$, then define the potential with the inverse Fourier transformation of the $M(\boldsymbol{k})/K^2$ . The reason for choosing  $D_i=\sinh{2\pi i k_i/n}$ is that we compute the derivative of the potential in this way $\partial_x\Phi(i)\approx \frac{1}{2\Delta X}(\Phi(i+1)-\Phi(i-1))$.

At last, we calculate $\partial_i\partial_j\Phi$ at each grid and find $\hat{\boldsymbol{e}}_2$. Then for each non-empty grid, we sum $\hat{L}_i\hat{L}_j$  of each halo to form the matrix $A_{ij}=\braket{\hat{L}_i\hat{L}_j}$ in which $\boldsymbol{\hat{L}}$ is a unit vector of halo angular momentum. Now, we find the eigenvector of $A_{ij}$ associated with the largest eigenvalue. We call this vector $\boldsymbol{\hat{J}}$ that is the direction most halos tend to align. Now, we calculate the angle between $\boldsymbol{\hat{J}}$  and $\hat{\boldsymbol{e}}_2$ and call it $\alpha$. If TTT were the only mechanism to explain the angular momentum, then one must have $|\cos\alpha| \sim 1$. We will check this in {Sec. \ref{sec:results}}. One could argue that we could find the angle between the spin vector of each halo and the $\hat{\boldsymbol{e}}_2$ at each grid.

{Note that in order to find the tidal tensor, we smoothed the potential on a scale of $10\mathrm{Mpc}$ while each halo experiences the local tidal tensor at its position, which perhaps differs from the average tidal tensor referred to of the whole grid.  In this sense,  halos, on average, are aligned with $\hat{\boldsymbol{e}}_2$.  As a result, at each grid we first find $\boldsymbol{\hat{J}}$, and check for the excess probability of $\boldsymbol{\hat{J}}.\hat{\boldsymbol{e}}_2$.}

\begin{figure}
\centering
    \includegraphics[scale=0.4]{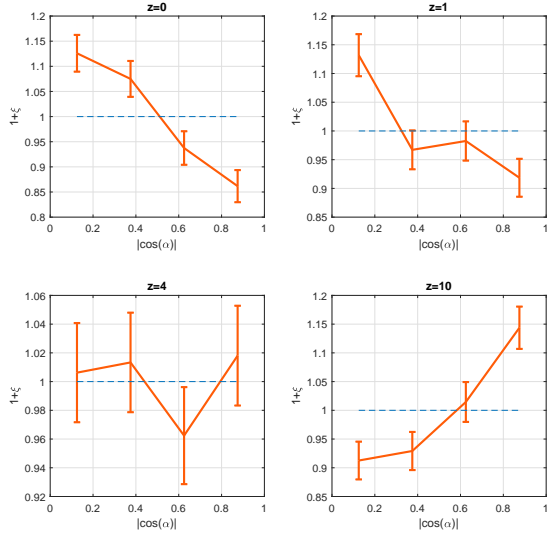}
    \caption{Spin alignment for TTT. $\alpha$ is the angle between $\boldsymbol{\hat{e}}_2$ and $\boldsymbol{\hat{J}}$}
    \label{fig:LvsPhi}
\end{figure}

\section{Results and Conclusion}
\label{sec:results}
We know that halos' angular momentum can be described solely by the TTT at the early stages of structure formation. However, as time goes by and fly-by events become frequent, the TTT's prediction for the alignments is no longer valid. On the other hand, if not impossible, it is cumbersome to predict angular momentum alignment of an arbitrary halo during non-linear stages analytically. Nevertheless, the TLT predicts that the spin of the large mad halos is likely to align with their speed. The reason is that they experience the tidal locking effect more than other halos. The alignment of the mad halos must be most substantial at late times while it is absent at the higher redshifts.

\begin{figure}
\centering
    \includegraphics[scale=0.4]{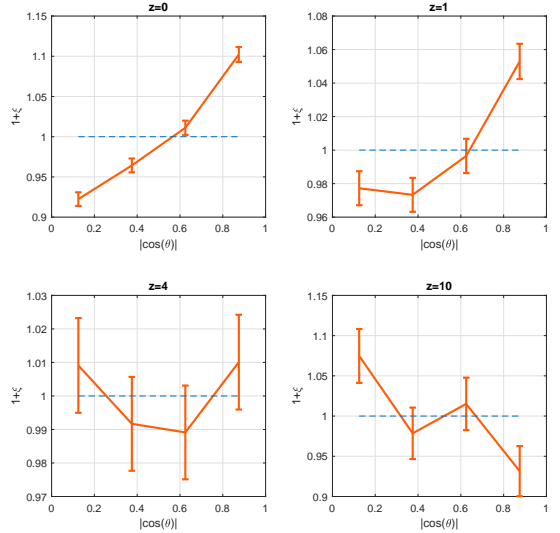}
    \caption{Spin alignment for large mad-halos. $\theta$ is the angle between speed and angular momentum.}
    \label{fig:dilutMadH1}
\end{figure}

First, let us start with checking the TTT predictions versus simulation. The Fig \ref{fig:LvsPhi} is to examine TTT. As already expected, the orientation of the halos' angular momentum agrees with TTT predictions at high redshifts while it diminishes later on. It shows an excess probability of alignment up to 15 percent at $z=10$, while there is no excess probability even at $z=4$. We can conclude that TTT is responsible for the initial angular momentum of the halos. However, subsequently, something changes the halos' angular momentum and eliminates the predicted alignment.

\begin{figure}
\centering
        \includegraphics[scale=0.4]{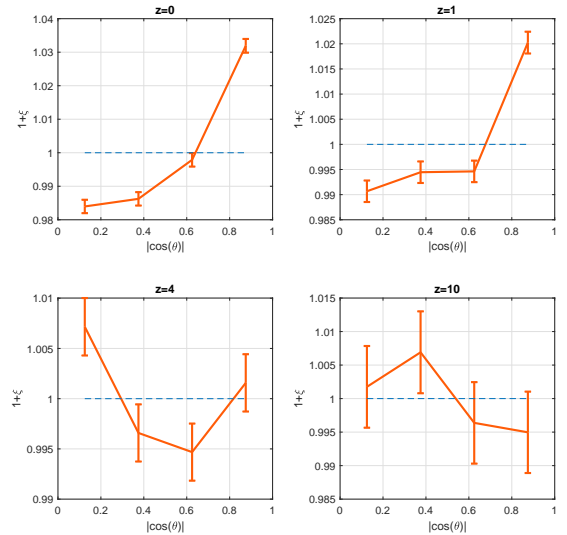}
    \caption{Spin alignment for mad-halos which are not necessarily large. $\theta$ is the angle between speed and angular momentum.}
    \label{fig:MadH1}
\end{figure}

We claim that the tidal locking mechanism changes the angular momentum of the halos, and it causes the alignment between spin and speed for large mad halos. Fig. \ref{fig:dilutMadH1} shows  $\cos\theta$ histogram in different redshifts for the large mad halos, where $\theta$ is the angle between speed and the angular momentum. As shown in Fig.1, the halos’ angular momenta show \emph{partial alignment} with their velocity that is statistically significant. At $z=0$ the excess probability of alignment is about 10 percent, then reduces at $z=1$ to 5 percent and completely disappears at the redshift $z=4$ and $z=10$.

Moreover, as discussed earlier,  large halos would be more significantly affected by the TLT effect, so we should see a reduction in alignment if we relax the radius constraint. Fig. \ref{fig:MadH1} shows the $\cos\theta$  distribution for all mad halos, not necessarily the large ones. The excess probability reduces by a factor of $3$ and disappears at earlier redshifts.

\begin{figure}
\centering
    \includegraphics[scale=0.4]{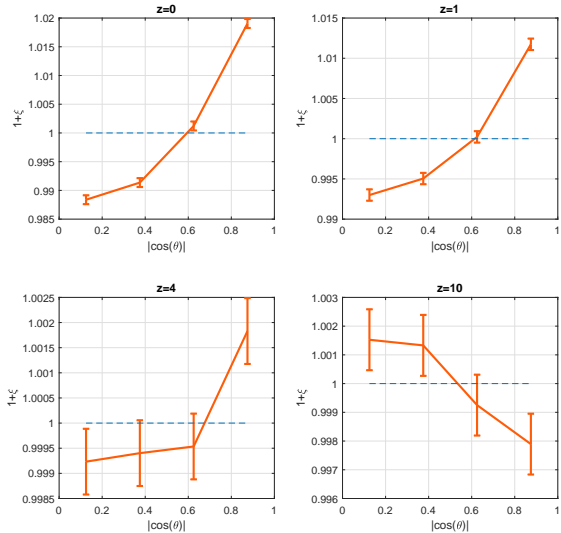}
    \caption{Spin-speed alignment for all of the halos. $\theta$ is the angle between speed and angular momentum.}
    \label{fig:LvsV}
\end{figure}

{On the other hand, if we relax the velocity constraint, we still see partial spin-speed alignment between all halos in the simulation.  Fig.\ref{fig:LvsV} shows that while it is statistically significant, the excess probability of alignment is less than 2 percent if we include all halos regardless of the velocity.  As we expect, alignment completely disappears at $z=10$, so one can say that the partial alignment developed only during non-linear stages, and perhaps it is due to TLT.}

{The above findings confirm the TLT theory. Once the dark matter halo forms, its spin direction follows the TTT prediction; however, when enough time goes by, mergers and the TLT changes the angular momentum orientation. Some studies focused on the mass dependence of spin-LSS alignments. However, TLT predictions show that spin-LSS, on the fundamental level, depends on the size and speed of the halo. For example, since the halos' speed in filaments is anisotropic, one may expect to see a non-trivial spin alignment. Studying filaments in phase space and taking TLT into account will help us to understand observed spin-LSS alignment.
\\
We must mention that linking these results to the observations is not straightforward. We can only measure the universe's baryonic matter density, so it is not easy to estimate the galaxies' angular momenta solely by knowing the spin of the host dark matter halo. However, the IllustrisTNG simulation has baryonic matter, and one can use it to address this question. It is indeed an interesting question that will be studied thoroughly in independent work.
}

\section*{Acknowledgment}
We would like to thank Shant Baghram, Mohammad Ansarifard, Sohrab Rahvar, Pablo López, and Corentin Cadiou for their fruitfull comments and discussions.

\bibliography{Mad_Halos.bib}{}
\bibliographystyle{aasjournal}



\end{document}